\documentclass[lettersize,journal]{IEEEtran}
\usepackage{amsmath,amsfonts}
\usepackage{algorithm}
\usepackage{algpseudocode}
\usepackage{array}
\usepackage[caption=false,font=normalsize,labelfont=sf,textfont=sf]{subfig}
\usepackage{textcomp}
\usepackage{stfloats}
\usepackage{url}
\usepackage{graphicx}
\usepackage{verbatim}
\usepackage{cite}
\usepackage{svg}
\usepackage{threeparttable}
\usepackage{listings}
\usepackage{xcolor}
\usepackage{amssymb}
\usepackage{booktabs}
\usepackage[most]{tcolorbox}
\begin{document}

\title{Analysis of LLM Vulnerability to GPU Soft Errors: An Instruction-Level Fault Injection Study}

\author{
    Duo Chai,
    Zizhen Liu,~\IEEEmembership{Member,~IEEE},
    Shuhuai Wang,
    Songwei Pei,~\IEEEmembership{Member,~IEEE},
    Cheng Liu,~\IEEEmembership{Senior Member,~IEEE},
    Huawei Li,~\IEEEmembership{Senior Member,~IEEE},
    and Shangguang Wang,~\IEEEmembership{Senior Member,~IEEE}
    \thanks{
    This paper is supported in part by the Shandong Provincial Natural Science Foundation (ZR2024LZH010), the Science and Technology Innovation Program of Xiongan New Area (2024XAGG0002), the National Natural Science Foundation of China (NSFC) under grant No. 62174162 and No. 62090024.}
    \thanks{Duo Chai, Shuhuai Wang, Songwei Pei and Shangguang Wang are with the School of Computer Science (National Pilot Software Engineering School), Beijing University of Posts and Telecommunications, Beijing 100876, China. 
    Zizhen Liu and Huawei Li are with the State Key Laboratory of Processors, Institute of Computing Technology, Chinese Academy of Sciences, Beijing 100190, China.
    Cheng Liu is with the School of Control and Computer Engineering, North China Electric Power University, Beijing 102206, China.
    (e-mail: chaiduo@bupt.edu.cn, liuzizhen@ict.ac.cn, wangshuhuai@bupt.edu.cn, peisongwei@bupt.edu.cn, liucheng@ncepu.edu.cn, lihuawei@ict.ac.cn, sgwang@bupt.edu.cn). (corresponding authors: Songwei Pei, Cheng Liu).}
}

\markboth{IEEE Transactions on Computer-Aided Design of Integrated Circuits and Systems,~Vol.~XX, No.~X, Month~Year}%}%
{Chai \MakeLowercase{\textit{et al.}}: Analysis of LLM Vulnerability to GPU Soft Errors}

% \IEEEpubid{0000--0000/00\$00.00~\copyright~2025 IEEE}

\maketitle

\begin{abstract}
Large language models (LLMs) are highly compute- and memory-intensive, posing significant demands on high-performance GPUs. At the same time, advances in GPU technology driven by shrinking transistor sizes and lower operating voltages have made these devices increasingly susceptible to soft errors, particularly in radiation-prone environments such as satellite computing scenarios. While prior work has examined GPU soft errors, most studies have focused on general-purpose applications or conventional neural networks primarily used for vision tasks such as classification and detection. However, vulnerability analysis of modern LLMs remains limited, despite their rapid adoption in diverse application scenarios. Given the unique characteristics of LLMs, their resilience to soft errors may differ substantially from earlier models. To bridge this gap, we conduct the first instruction-level fault injection study of LLM inference to analyze the factors affecting LLM vulnerability. Our approach reveals reliability characteristics from multiple perspectives, highlighting the effects of model architecture, parameter scale, and task complexity. These findings provide new insights into LLM reliability and inform the design of more effective fault tolerance mechanisms.
\end{abstract}


\section{Introduction}
\IEEEPARstart{I}{n} recent years, the rapid advancement of large language models (LLMs) such as GPT-4 \cite{achiam2023gpt} and DeepSeek \cite{guo2025deepseek} has marked a significant milestone in the pursuit of general artificial intelligence. Unlike earlier convolutional neural network models primarily designed for computer vision tasks, LLMs exhibit strong capabilities in natural language understanding, reasoning, and generation, reflecting both broader application scenarios and fundamentally different computational patterns. Executing LLM inference demands substantial computational resources and is widely accelerated by high-performance GPUs, supported by the CUDA programming ecosystem. However, modern GPUs are increasingly vulnerable to transient faults as transistor sizes and operating voltages continue to shrink, especially in satellite computing scenarios under high-radiation environments. Such faults, which may arise from temperature or voltage fluctuations or from high-energy particle strikes in the environment \cite{kalra2020armorall}, can corrupt logic or storage states, most commonly manifesting as single-bit flips \cite{dodd2003basic}. Once generated, these soft errors may propagate through the computation graph of an LLM, leading to different inference outcomes.
\begin{figure}[t]
    \centering
    \includegraphics[width=0.46\textwidth]{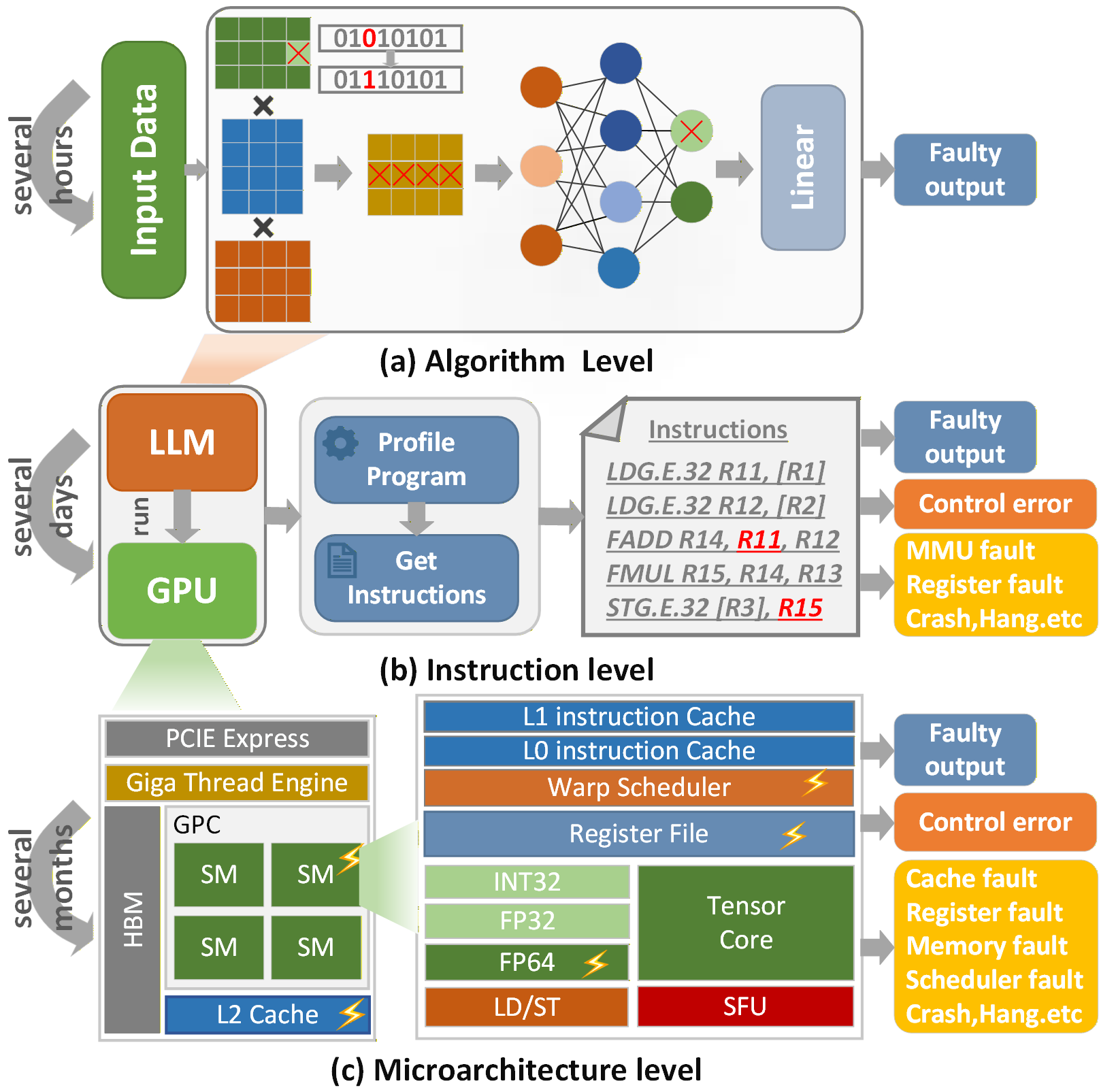}
    \caption{Typical fault injection approaches at different abstraction levels. (a) Algorithm-level fault injection, which injects bit-flip errors into weights and activations at neural network layers. (b) Instruction-level fault injection, which injects bit-flip errors into GPU instructions. (c) Microarchitecture-level fault injection, which injects bit-flip errors at major registers and components of the GPU. Essentially, microarchitecture-level fault injection depicts how hardware errors propagate to instructions and algorithms eventually. It is accurate but can be extremely slow. In contrast, algorithm-level fault injection simplifies the error modeling to network operations and ignores low-level details for higher fault simulation speed. Instruction-level fault injection stays between microarchitecture level and algorithm-level fault injection for balanced fault simulation accuracy and speed.}
    \vspace{-2em}
    \label{fig1}
\end{figure}
The consequences of LLMs under GPU soft errors can be categorized into three types~\cite{guerrero2023understanding,roquet2024cross,nie2018fault}: (1) Masked, where faults are successfully masked and model inference remains unaffected; (2) SDC (Silent Data Corruption), where the model produces incorrect results without any system warning (in this work, we adopt the standard definition of SDC, where any token-level deviation from the fault-free output is classified as an SDC); and (3) DUE (Detected Unrecoverable Error), where unrecoverable errors cause program termination or crashes. LLMs are increasingly being deployed in safety-critical domains such as autonomous driving and intelligent healthcare, where even minor soft errors may lead to catastrophic consequences. Therefore, a comprehensive understanding of the impact of GPU soft errors on LLMs is urgently needed.

% \IEEEpubidadjcol
Significant efforts have been devoted to studying GPU vulnerability to soft errors and mitigating their impact on AI models. However, most existing fault-injection studies for neural networks focus on coarse-grained perturbations, such as weight corruption or layer-wise activation bit-flips as shown in Figure~\ref{fig1}(a). While these approaches are efficient and provide useful insights, they fail to capture the fine-grained computational characteristics of LLMs—such as KV-cache optimizations and operator fusion—which are naturally suited for instruction-level fault injection. Some prior work~\cite{dos2021revealing,esposito2024evaluating} has investigated the vulnerability of convolutional neural networks (CNNs) to GPU soft errors using instruction-level fault injection, but their conclusions may not generalize to LLMs, which have fundamentally different architectures and tasks. CNNs typically rely on convolutional operators and target vision tasks such as classification and detection, whereas LLMs are based on Transformer architectures and address a broader set of tasks, including reasoning and generation. In addition, the model accuracy metrics also vary substantially. CNN-based vision tasks generally have specific accuracy metrics, while LLMs need more complex metrics for evaluation, such as Perplexity (PPL) and Exact Match, due to the inherent task complexity. There are also several studies \cite{dos2021revealing, yang2024gpu, bolchini2023analyzing, sartzetakis2022gpufi, guerrero2023understanding, van2024improving,tan2025gerem} analyzing programs on CPUs or GPUs at the RTL or microarchitecture level using simulators as shown in Figure~\ref{fig1}(c). However, such approaches are extremely time-consuming, especially for compute-intensive LLMs, and existing simulators may not accurately capture the behavior of contemporary GPU architectures.

To achieve both efficient and accurate fault injection evaluation for LLM vulnerability, we adopt NVBitFI \cite{tsai2021nvbitfi}, an instruction-level fault injection framework (Figure~\ref{fig1}(b)), to comprehensively investigate the impact of GPU soft errors in LLMs. We first characterize the distribution of abnormal behavior in LLM inference systems and analyze its underlying causes. Moreover, we perform a systematic study of how instruction types, bit positions, component types, layer variations, and task difficulty influence LLM reliability across a suite of representative tasks, including reasoning and text summarization. Our results demonstrate that task difficulty, model scale, and architectural design are key determinants of model vulnerability. To the best of our knowledge, this work is the first to establish a direct connection between GPU soft errors and high-level LLM vulnerability through instruction-level fault injection, providing new insights into the robustness of LLM inference systems.

The main contributions of this work are as follows:
\begin{enumerate}
\item This is the first attempt to systematically analyze the vulnerability of LLMs on GPUs with instruction-level fault injection. Particularly, we analyze the abnormal behaviors of LLM inference in the presence of soft errors and identify their underlying causes. 

\item We quantify the vulnerability factors of key GPU instruction types and derive an efficient overall vulnerability metric for LLMs based on these factors. 

\item We perform a comprehensive vulnerability analysis of LLMs from five perspectives, including instruction types, bit positions, fault layers, component types and task difficulty, revealing how critical architectural parameters of LLMs influence their robustness against soft errors.
\end{enumerate}

\section{Related Work}
\subsection{Fault Injection}
Fault injection is a common method for evaluating program reliability~\cite{van2024improving, wei2014quantifying,he2023demystifying, wang2023understanding,huang2024mrfi}. Currently, based on the abstraction levels of fault injection, it can be roughly classified into three categories: microarchitecture-level, instruction-level, and algorithm-level fault injection. Microarchitecture-level and instruction-level fault injection can effectively reflect the hardware characteristics of GPUs, whereas algorithm-level fault injection cannot.

Microarchitecture-level fault injection \cite{yang2024gpu, tan2025gerem, condia2021combining} is typically implemented using the GPGPU-Sim simulator \cite{bakhoda2009analyzing}, a cycle-accurate and trace-driven simulator that provides a detailed model of GPU microarchitectures. However, the simulation speed can be orders of magnitude slower than on realistic GPUs because the low-level simulation involves too many details. This becomes a major barrier in evaluating complex applications such as LLMs, which typically require a substantial amount of execution time \cite{avalos2021principal}. In addition, the microarchitectures of the state-of-the-art (SOTA) GPUs are usually unknown to the public, making the fault simulation limited to only old GPU architectures, which can be distinct from SOTA GPUs. 

Instruction-level fault injection, on the other hand, is usually performed using specialized tools such as GPU-Qin~\cite{fang2014gpu},
LLFI-GPU~\cite{li2016understanding}, SASSFI~\cite{hari2017sassifi}, and NVBitFI~\cite{tsai2021nvbitfi}, which inject bit-flip errors into GPU instructions. Since the GPU instruction set is usually available and the fault injection can also be applied on recent off-the-shelf GPU architectures, the simulation speed can be much faster than that of the microarchitecture-level fault injection tools and enables evaluation of large-scale applications on modern GPU architectures. It has been widely adopted for reliability studies on general GPU programs \cite{dos2021revealing, yang2024gpu} and CNNs \cite{bolchini2023analyzing,condia2021combining, huang2023statistical,xu2021reliability,xue2023soft,guerrero2022effective,li2017understanding,he2023understanding} with a lack of work specifically targeting LLMs. 

Algorithm-level fault injection primarily injects faults into model weights or activation values to perturb the model inference process. These fault injection tools typically offer high execution speed. Representative works include PyTorchFI~\cite{mahmoud2020pytorchfi}, which leverages PyTorch’s hook mechanism to inject bit-flip errors during the model’s forward propagation. Other notable examples include MRFI~\cite{huang2024mrfi} and TensorFI~\cite{chen2020tensorfi}. Since these tools are generally independent of the underlying computing engines, they fail to capture GPU-specific characteristics. As a result, they are primarily limited to investigating computational errors and are mostly employed for algorithm-level fault-tolerant design studies.

\subsection{Reliability Analysis of AI Models}
Artificial intelligence (AI) has been increasingly adopted in safety-critical applications such as autonomous driving and robotics, where hardware errors can lead to catastrophic consequences. Consequently, extensive research has been conducted on the reliability of AI models over the past decade \cite{dos2021revealing, bolchini2023analyzing, condia2021combining, li2017understanding, he2023understanding,liu2022special, he2020fidelity, hsiao2023silent,li2023built}.
Existing studies on AI model reliability can be roughly categorized into two directions.

The first category \cite{ma2024dr,roquet2024cross,he2023understanding,ma2025understanding} investigates the impact of hardware faults on AI model behavior and performance metrics, such as accuracy degradation and model vulnerability. These studies aim to understand how hardware errors propagate through neural networks and affect overall model robustness. For example, Roquet et al.~\cite{roquet2024cross} systematically evaluated the runtime reliability of large Vision Transformer (ViT) models across two NVIDIA GPU architectures, while He et al.~\cite{he2023understanding} characterized the effects of hardware faults on neural network training processes. Ma et al.~\cite{ma2024dr} proposed an SDC detection and mitigation framework based on neuron activation distribution.

The second category \cite{bolchini2023analyzing,li2017understanding,ma2023error,xie2025realm,agarwal2023resilience} explores design factors that influence AI model reliability—including quantization levels, bit positions, convolution implementations, and layer characteristics—which can provide valuable insights for fault-tolerant AI model design. For instance, Bolchini et al.~\cite{bolchini2023analyzing} compared the reliability of neural networks implemented with GEMM, FFT, and Winograd convolution methods. Li et al.~\cite{li2017understanding} conducted an in-depth analysis of how data types, bit positions, layer properties, and data reuse patterns affect error propagation in neural networks. With the emergence of transformer-based architectures, researchers have also begun examining their reliability. Ma et al.~\cite{ma2023error} analyzed soft error vulnerability in various transformer modules used in ViTs, revealing that certain layers are particularly sensitive to single-bit flips. Similarly, Xie et al.~\cite{xie2025realm} systematically characterized fault tolerance across different layers, bit positions, and architectural components, and proposed a circuit-algorithm co-design fault-tolerance framework. 

Despite these advances, existing reliability studies still exhibit several limitations: 
(1) Most prior work focuses on CNNs, and their conclusions may not directly generalize to transformer-based architectures. This is because LLMs perform fundamentally different tasks, such as reasoning and summarization, compared to traditional vision tasks, and they also rely on different evaluation metrics. 
(2) Although recent efforts have begun to explore the reliability of transformers, most of them remain at algorithm-level fault injection, which only reflects abstract error characteristics and fails to accurately capture real GPU execution behavior.

\section{LLM Reliability Evaluation Framework}

\begin{table}[htbp]
\caption{Comparison between our framework and NVBitFI ($k$ denotes the number of faults per run).
\label{table1}}
\centering
\renewcommand{\arraystretch}{1.05}
\begin{tabular}{c c c}
\toprule
Ability & NVBitFI & Ours \\
\midrule
Faults per run & 1 & $k$ \\
Bit flip mode & random & random \& fixed \\
Instruction selection  & random & random \& fixed \\
LLM-based evaluation & unsupported & supported \\
\bottomrule
\end{tabular}
\end{table}

\begin{figure*}[htbp]
    \centering
    \includegraphics[width=0.96\textwidth]{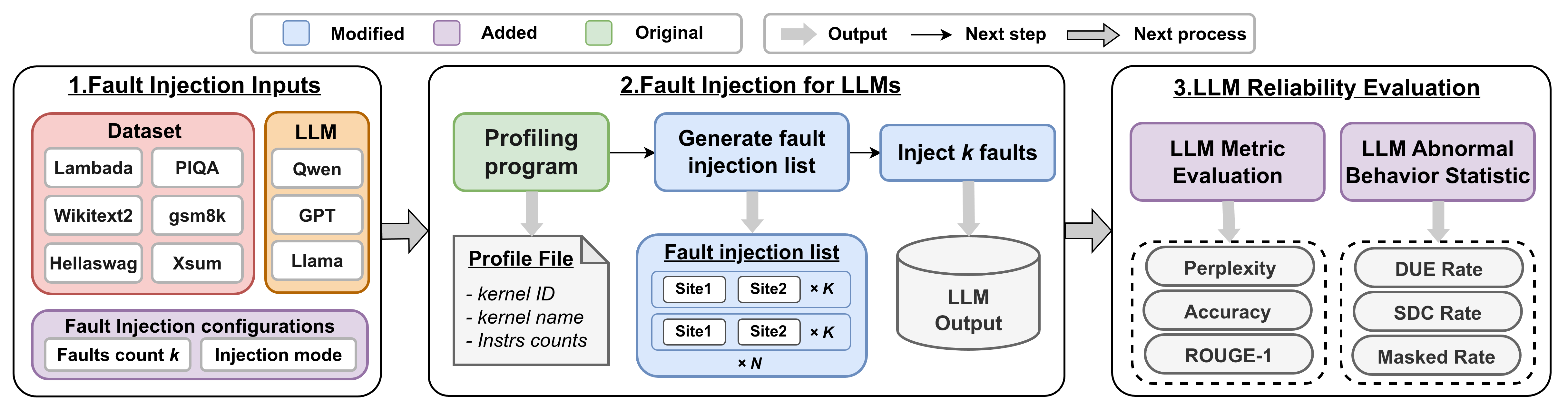}
    \caption{The proposed LLM reliability evaluation framework based on NVBitFI.}
    \label{fig2}
\end{figure*}

To address the above limitations, we propose an NVBitFI-based reliability evaluation framework for LLMs that enables comprehensive, instruction-level fault injection on modern GPU devices. This framework facilitates a more accurate and practical analysis of GPU soft errors, providing insights for developing fault-tolerant LLMs in real-world GPU environments.

Table~\ref{table1} presents a functional comparison between our proposed framework and NVBitFI. Although NVBitFI~\cite{tsai2021nvbitfi} provides a convenient mechanism for instruction-level fault injection in generic CUDA programs, considerable effort is still required to extend it for LLM reliability evaluation under diverse metrics and fault injection configurations. Figure~\ref{fig2} illustrates the proposed LLM reliability evaluation framework built upon NVBitFI. The framework comprises three main components: (1) fault injection inputs; (2) fault injection for LLMs; (3) LLM reliability evaluation. 

\subsection{Fault Injection Inputs}
Fault injection inputs include the target model, the evaluation dataset, and the fault injection configurations. Specifically, users can specify the number of injected faults and select different injection modes, such as random injection, fixed instruction-type injection, or fixed bit-position injection. To support LLM-oriented reliability studies, we extend the primitive bit-flip mechanism in NVBitFI with more flexible injection modes. These extensions allow faults to be injected into selected GPU instruction types and enable the injected faults to be further associated with LLM layers or modules based on runtime execution metadata and output behaviors. This design makes it possible to analyze how GPU soft errors propagate to high-level LLM components, which is essential for understanding the vulnerability characteristics of LLM inference.

\begin{algorithm}[t]
\caption{Multi-Fault List Generation}
\label{alg:faultlist}
\begin{algorithmic}[1]
\Require \\
Number of injection runs $N$ \\
Fault number per run $k$ \\
Instruction mode $I\in\{\textit{Random},\textit{Fixed}\}$ \\
Bit mode $B\in\{\textit{Random},\textit{Fixed}\}$ \\
Target instruction type $T$ if $I=\textit{Fixed}$ \\
Target bit position $P$ if $B=\textit{Fixed}$
\Ensure Fault injection list $F$
\State $F\gets\emptyset$
\For{$r=1$ to $N$}
    \State $E\gets\emptyset$
    \While{$|E|<k$}
        \State $(kid,kernel)\gets$ SelectKernelInstance()
        \If{$I=$ \textit{Random}}
            \State $inst\gets$ RandomInst($kernel$)
        \Else
            \State $inst\gets$ SelectInstByType($kernel,T$)
        \EndIf
        \State $register\gets$ SelectRegister($inst$)
        \If{$B=$ \textit{Random}}
            \State $bit\gets$ RandomBit($register$)
        \Else
            \State $bit\gets P$
        \EndIf
        \State $f \gets \langle kid,kernel,inst,register,bit,\textit{False}\rangle$
        \If{$f\notin E$}
            \State $E \gets E \cup \{f\}$
        \EndIf
    \EndWhile
    \State $F\gets F\cup\{E\}$
\EndFor
\State \Return $F$
\end{algorithmic}
\end{algorithm}

\begin{algorithm}[t]
\caption{Runtime Multi-Fault Injection}
\label{alg:inject}
\begin{algorithmic}[1]
\Require Fault entry $E=\{f_1,\ldots,f_k\}$
\State $InjectedCnt\gets0$
\Procedure{InstructionCallback}{$kid,kernel,inst$}
\ForAll{$f_i\in E$}
\If{Match($kid,kernel,inst,f_i$)}
\If{AtomicCAS($f_i.injected$, False, True)}
\State InjectFault($f_i.register,f_i.bit$)
\State AtomicAdd($InjectedCnt,1$)
\EndIf
\EndIf
\EndFor
\EndProcedure
\State Launch the target application
\While{the target application is running}
\State Wait for callbacks triggered by instrumented instructions
\EndWhile
\State Record the status according to $InjectedCnt=|E|$
\end{algorithmic}
\end{algorithm}

\subsection{Fault Injection for LLMs}
The framework leverages NVBit~\cite{villa2019nvbit}, the dynamic binary instrumentation library on which NVBitFI is built, to intercept GPU kernel operations, collect instruction characteristics, and identify valid fault injection locations. To support the injection of multiple faults within a single execution run, we modify the fault list generation code such that each entry in the generated fault injection list contains multiple fault locations, as shown in Algorithm~\ref{alg:faultlist}. In this algorithm, \texttt{SelectKernelInstance()} samples a kernel instance recorded during a profiling run; \texttt{RandomInst()}/\texttt{SelectInstByType()} selects an injectable instruction (one with a register operand) uniformly or by opcode $T$; and \texttt{SelectRegister()} picks a source-register operand, where the target bit is selected randomly or fixed at position $P$ depending on $B$. These fault locations are subsequently used to guide the runtime fault injection process. We further modify the fault injection kernel to enable multiple faults to be injected in one execution, as described in Algorithm~\ref{alg:inject}. At runtime, the callback matches each executed instruction against pending faults, and \texttt{InjectFault()} flips the target bit of the selected register in the thread context. An atomic compare-and-swap on the \textit{injected} flag guarantees that each fault is applied exactly once, and the framework verifies $\mathit{InjectedCnt}=|E|$ per run. These capabilities are crucial for studying the vulnerability of LLMs under multi-fault scenarios.

\subsection{LLM Reliability Evaluation}
After fault injection, the framework collects LLM inference outputs. It then performs statistical analysis of abnormal behaviors and computes evaluation metrics such as accuracy, perplexity, and ROUGE-1. Through this process, the framework enables comprehensive, high-level reliability evaluation of LLM systems.

\section{Basic Experimental Setup}
\subsection{Experimental Environment}
All experiments in this study were conducted using the proposed framework on a server equipped with an NVIDIA A100 GPU (80 GB), an Intel Xeon Platinum 8358P CPU, and 512 GB of system memory. The software environment consisted of CUDA 12.2, Python 3.12, and NVBit 1.7.5. 

\subsection{Models and Datasets}
We selected three representative LLM architectures: GPT2 \cite{radford2019language}, Llama3.2 \cite{meta2024llama32}, and Qwen3 \cite{qwen3technicalreport}. Their detailed configurations are summarized in Table~\ref{table2}. For each architecture, two model scales were evaluated. 
To comprehensively assess model performance, we adopted benchmark datasets including Lambada \cite{paperno2016lambada}, PIQA \cite{bisk2020piqa}, HellaSwag \cite{zellers2019hellaswag}, WikiText-2 \cite{merity2016pointer}, XSum \cite{narayan2018don}, and GSM8K \cite{cobbe2021training}, covering a diverse range of tasks including text generation, mathematical reasoning, and language modeling.

\subsection{Fault Injection Experiments}
All fault injection locations were selected through random sampling over the dynamic instruction stream collected during model execution. For multi-fault scenarios, each fault was sampled independently. A summary of the fault injection experiments presented in this paper is provided in Table~\ref{table3}.

\begin{table}[t]
\caption{Detailed parameter description of the selected models.
\label{table2}}
\renewcommand{\arraystretch}{1.05}
\centering
\begin{tabular}{c c c c c c c}
\toprule
Model & P & L & HS & AH & VS & CL\\
\midrule
GPT2 & 124M & 12 & 768 & 12 & 50257 & 1k\\
GPT2-large & 774M & 36 & 1280 & 20 & 50257 & 1k\\
Llama3.2-1B  & 1.23B & 16 & 2048 & 32 & 128256 & 128k\\
Llama3.2-3B & 3.21B & 28 & 3072 & 24 & 128256 & 128k\\
Qwen3-0.6B & 0.59B & 28 & 1024 & 16 & 151936 & 40k\\
Qwen3-1.7B & 1.72B & 28 & 2048 & 16 & 151936 & 40k\\
\bottomrule
\end{tabular}
\begin{tablenotes}
\footnotesize
\item \textbf{Notes:} P=Parameters, L=Layers, HS=Hidden size, AH=Attention heads, VS=Vocabulary Size, CL=Context Length. 1k=1024, 1M=1 Million, 1B=1 Billion.
\end{tablenotes}
\end{table}

\begin{table}[t]
\centering
\caption{Summary of Fault Injection Experiments.}
\label{table3}
\renewcommand{\arraystretch}{1.05}
\begin{tabular}{l c c l}
\toprule
\textbf{Experiment} & \textbf{Fault Mode} & \textbf{Fault Count} & \textbf{Run Count} \\
\midrule
Abnormal behavior & Random & 1--8 & 4,000 per model \\
DUE causes & Random & 1 & 3,000 per model \\
SDC causes & Random & 1 & 3,000 per model \\
Instruction types & Fixed type & 1 & 500 per type \\
Bit positions & Fixed bit & 1 & 500 per bit \\
Component types & Random & 1 & 5,000 per model \\
Layers & Random & 1 & 5,000 per model \\
Tasks & Random & 1--8 & 4,000 per model \\
\bottomrule
\end{tabular}
\end{table}

\begin{figure*}[htbp]
    \centering
    \includegraphics[width=0.96\textwidth]{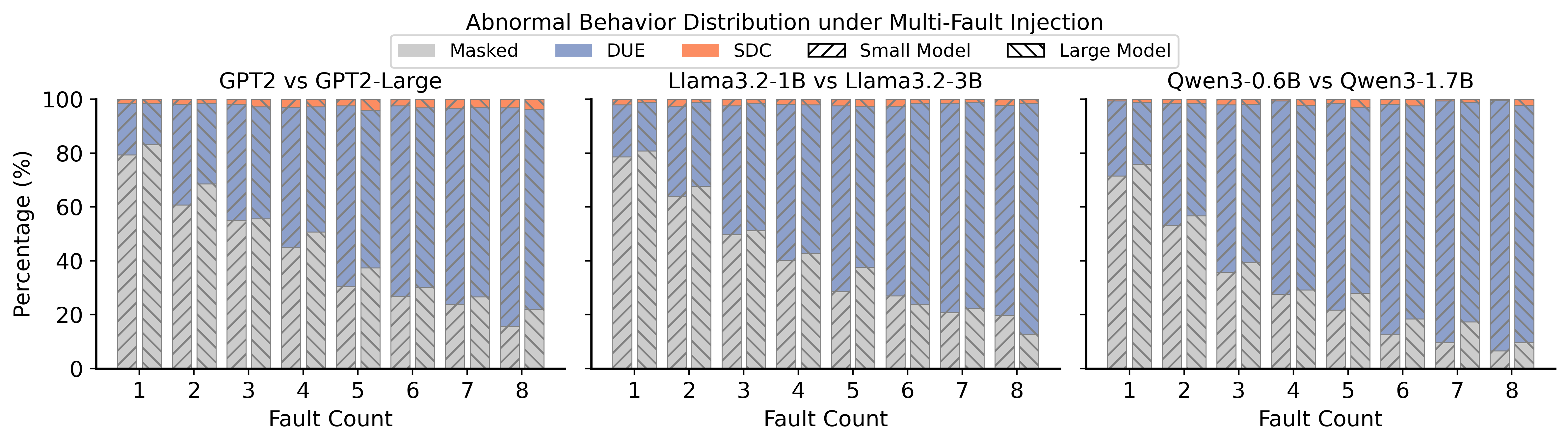}
    \caption{GPU behavior distribution of the LLM inference system under multi-fault scenarios.}
    \label{fig3}
\end{figure*}

\section{Abnormal Behavior Analysis of LLMs}
This section investigates the impact of soft errors on LLM inference and evaluates the abnormal behavior of LLMs. We further categorize the DUE events observed during fault injection and analyze the instruction-level causes of SDCs.

\subsection{Overview of LLM Abnormal Behavior}

Unlike reliability studies based on algorithm-level fault injection, which primarily target data errors during computation, instruction-level fault injection enables a more comprehensive investigation of abnormal behaviors induced by GPU soft errors. Therefore, we begin with an analysis of the abnormal behaviors observed in the LLM system.

Following the experimental setup described in Section IV, we conducted random fault injection experiments. By performing 4,000 fault injection runs on each model, we obtained the results shown in Figure~\ref{fig3}. The proportions of Masked, DUE, and SDC are the ratios of the occurrence counts of the corresponding events to the total number of experiments, and the abnormal rate is the sum of the DUE and SDC proportions. As illustrated in this figure, abnormal behaviors vary substantially as the number of injected faults increases. Overall, the rate of abnormal outcomes rises sharply, mainly due to surges in both DUE and SDC. Specifically, averaged over all evaluated models, the abnormal rate increases from 21.7\% with a single injected fault to 85.6\% with eight injected faults, indicating severe degradation in inference correctness as bit-flip errors accumulate. Similar to conventional applications, most errors in LLMs can be masked when only a single bit flip is injected. Moreover, as the number of injected faults increases, the proportion of DUE rises rapidly, whereas SDC grows much more slowly and may even plateau. A possible explanation is that many potential SDCs are preempted by DUEs. Since SDCs are only counted when an inference completes with an incorrect result, an increasing number of fault-induced exceptions terminate executions prematurely, preventing these inferences from being classified as SDCs.

\subsection{The Causes of DUEs}

\begin{figure}[t]
    \centering
    \includegraphics[width=0.46\textwidth]{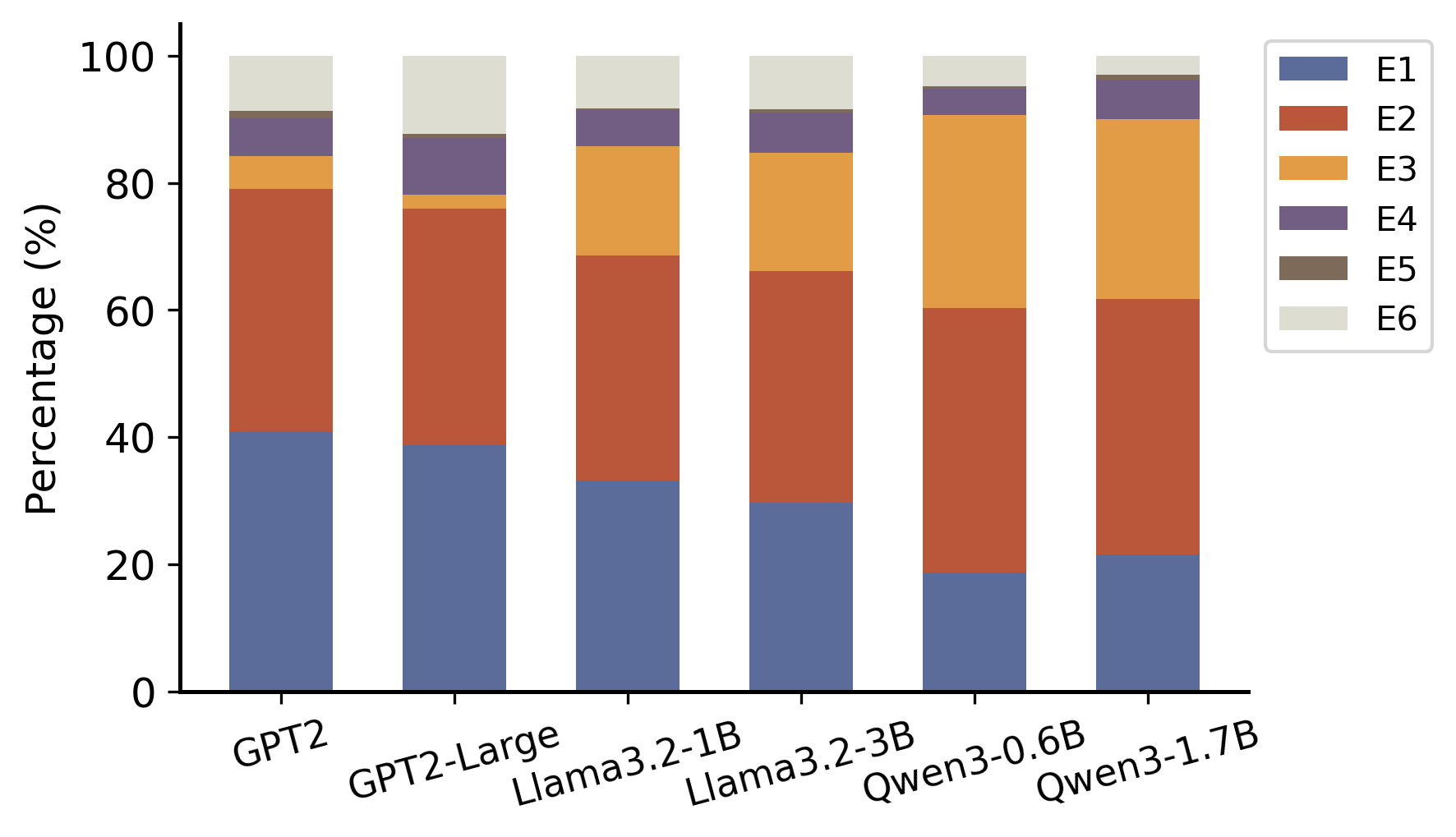}
    \caption{E1--E6 represent the following types of faults. E1: Out Of Range Address, E2: MMU Fault, E3: Out Of Range Register, E4: Misaligned Address, E5: Illegal Instruction Operands, E6: Others.}
    \label{fig4}
\end{figure}

Based on the experimental results shown in Figure~\ref{fig3}, we observe that even when only a single bit flip is injected, the proportion of DUE remains substantial. Moreover, as the number of injected faults increases, DUE gradually becomes the dominant type of anomalous behavior. To further investigate the root causes of DUE, we perform 3,000 fault injection experiments for each model and categorize the causes based on Xid error codes in the \texttt{dmesg} system log and CUDA runtime error messages, as summarized in Figure~\ref{fig4}.

The experimental results show that the causes of DUE can be broadly grouped into six categories: out of range address, MMU (Memory Management Unit) faults, out of range register, misaligned address, illegal instruction operands, and others. These categories are denoted as E1–E6, respectively. Among them, E1, E3, E4, and E5 are instruction-level errors detected by the CUDA runtime, most of which are related to illegal memory accesses; E2 corresponds to a hardware-level failure, in which the MMU itself raises a page-fault exception (reported as Xid 31) when a corrupted address fails page-table translation; E6 corresponds to temperature above threshold events, or other error conditions that cannot be unambiguously distinguished based on the available system logs. By summing the proportions of E1--E5 in Figure~\ref{fig4}, we observe that the attributable DUE errors exceed 90\%.

In our fault injection experiments, GPT2 is more prone to E1 errors, whereas Qwen and Llama more frequently trigger E3. One possible explanation lies in differences in model architectures and kernel implementations. GPT2 involves more memory accesses and address computations, which may increase the likelihood of memory-related faults and E1 errors. In contrast, Qwen and Llama employ highly fused, compute-intensive operators that place greater pressure on registers, making bit flips more likely to affect register-resident data and contribute to E3 errors.

\begin{figure}[t]
    \centering
    \includegraphics[width=0.44\textwidth]{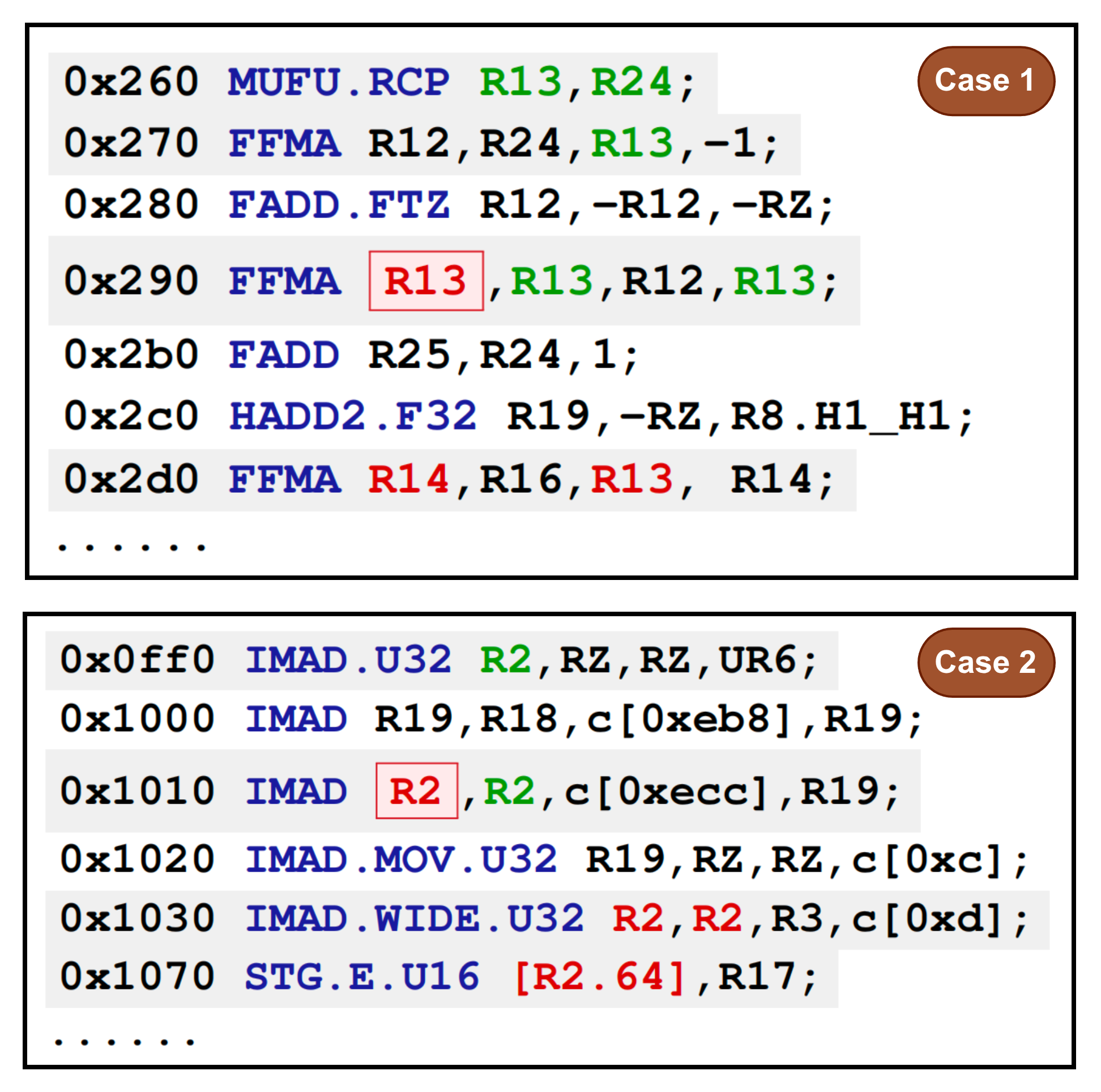}
    \caption{Case 1 illustrates an SDC caused by numerical errors. Case 2 illustrates an SDC caused by address errors. Green indicates registers containing correct data, while red indicates registers containing corrupted data, and the registers marked with red boxes represent registers where bit flip occurs.}
    \label{fig5}
\end{figure}

\subsection{The Causes of SDCs}
In this experiment, we performed 3,000 fault injection runs on each LLM, injecting a single fault in each run, and observed exactly 200 SDC instances. Through a detailed analysis of the corresponding Streaming Assembler (SASS) code for each SDC, we identified two primary causes: (1) numerical errors, particularly those that lead to large deviations, and (2) address errors that result in incorrect memory loading or writing. Among the observed cases, 89\% were attributed to numerical errors, while the remaining 11\% were due to address errors.

Large numerical deviations represent a major source of SDCs. A representative example is shown in Case 1 of Figure~\ref{fig5}. Register R13 was expected to be computed by the \texttt{MUFU.RCP} instruction as the reciprocal of R24, yielding a theoretical value less than 1. However, in the instruction \texttt{0x290 FFMA R13, R13, R12, R13}, a single bit flip altered R13’s floating-point value from \texttt{0x3e4ccccd} to \texttt{0x7e4ccccd}, causing the numerical result to jump from approximately 0.2 to about $6.8\times10^{37}$. This abnormal value subsequently propagated through later computations. For instance, in the instruction \texttt{0x2d0 FFMA R14, R16, R13, R14}, the value of R14 was excessively amplified, diverging completely from its expected range. As the fault continued to propagate, it ultimately resulted in an SDC.

Another example of SDC is shown in Case 2 of Figure~\ref{fig5}, which represents an address error. In the instruction \texttt{0x1010 IMAD R2, R2, c[0xecc], R19}, a bit flip occurred in R2, corrupting the stored address. This erroneous address was subsequently used in further calculations in \texttt{0x1030 IMAD.WIDE.U32 R2, R2, R3, c[0x160]}, and eventually in \texttt{0x1070 STG.E.U16 [R2.64], R17}. Since R2 was already corrupted, the correct value of R17 was written to an incorrect memory location, resulting in an SDC. In more severe cases, when the corrupted address becomes illegal, such as in out-of-bounds or misaligned memory accesses, it can lead to program crashes.

\section{Overall Vulnerability and Influencing-Factor Analysis of LLMs}
In this section, we first analyze the overall vulnerability of LLMs and quantify it using a vulnerability factor. We then propose an approximation algorithm to efficiently estimate model vulnerability and further validate the experimental findings.

Subsequently, based on the proposed framework, we conduct a comprehensive set of experiments to investigate the key factors affecting LLM vulnerability. Specifically, we organize our study according to a fault propagation hierarchy, covering three abstraction levels and five research dimensions:
\begin{enumerate}
\item{Instruction-Level}, including instruction types and bit positions, which determine where faults occur.
\item {Model-Level}, including component types and layer variations, which characterize how faults propagate through different model components (e.g., Attention) and layers.
\item {Task-Level}, which investigates how different tasks influence model vulnerability. 
\end{enumerate}

These three levels establish a hierarchical fault propagation path that bridges low-level instruction faults and high-level LLM behavior, providing a comprehensive understanding of model resilience under hardware-induced faults. Following the experimental setup presented in Section~IV, the experiments in this section took approximately 1,750 GPU hours.

\subsection{Overall Vulnerability Analysis of LLMs}

In this subsection, we first present the overall vulnerability of the evaluated LLMs and characterize their vulnerability based on the experimental results. We then propose an approximation algorithm that leverages instruction-level vulnerability to efficiently estimate model vulnerability, and use it to further corroborate the experimental findings.

\subsubsection{Model Vulnerability Factor}
\begin{figure}[htbp]
    \centering
    \includegraphics[width=0.46\textwidth]{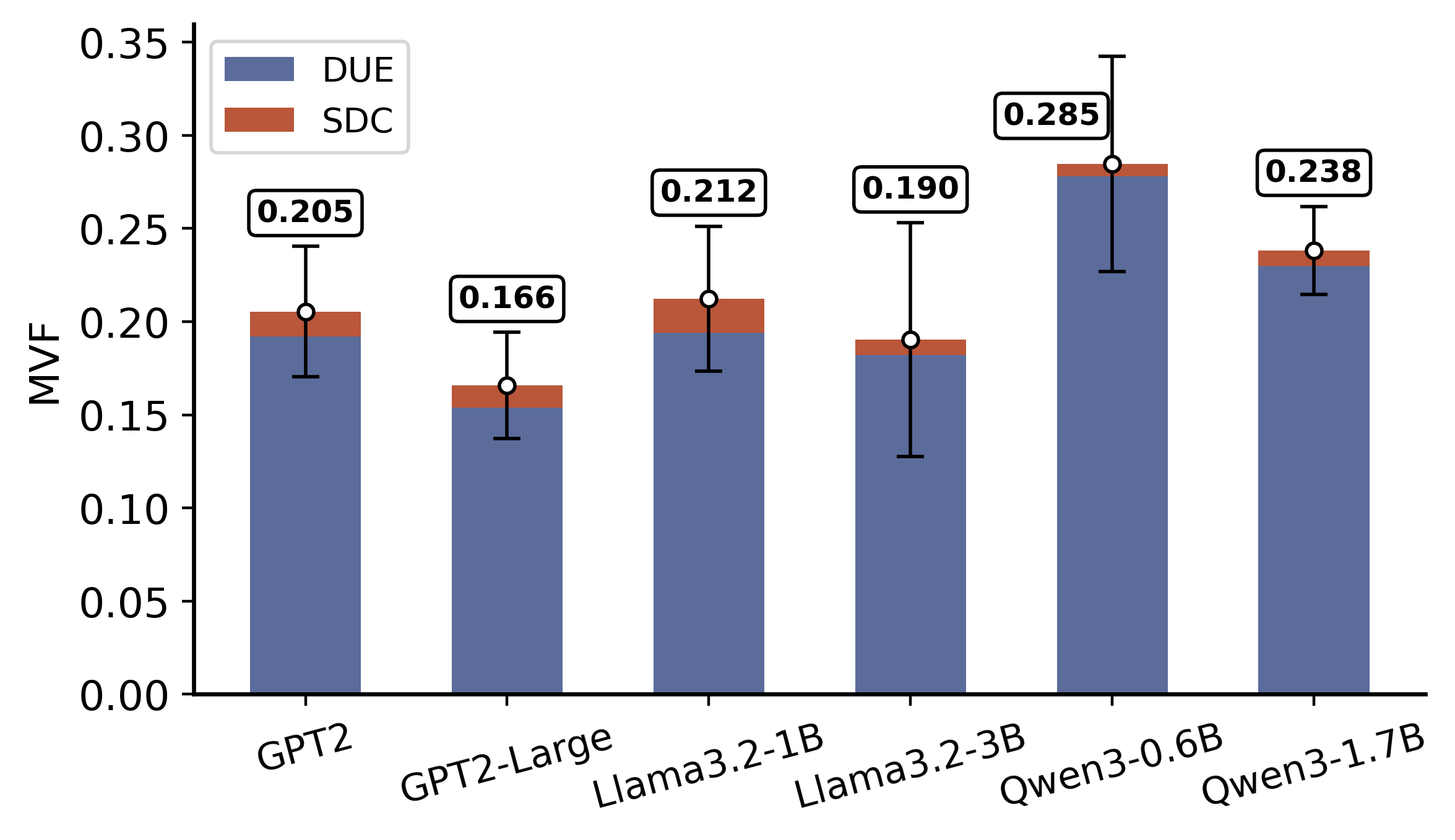}
    \caption{Model vulnerability factors of different LLMs. The error bars represent the standard error of the MVF over multiple experiments.}
    \label{fig6}
\end{figure}

To better compare the vulnerability of different models, we further derive the Model Vulnerability Factor (MVF), which quantifies the likelihood that a GPU soft error results in a visible failure in an LLM output, following the definition of the traditional program vulnerability factor. Essentially, the MVF is computed as the sum of the SDC rate and the DUE rate. As shown in Figure~\ref{fig6}, Qwen3-0.6B exhibits the highest MVF. Moreover, among models with the same architecture, larger models consistently achieve lower MVF than their smaller counterparts.

\begin{figure*}[htbp]
    \centering
    \includegraphics[width=0.96\textwidth]{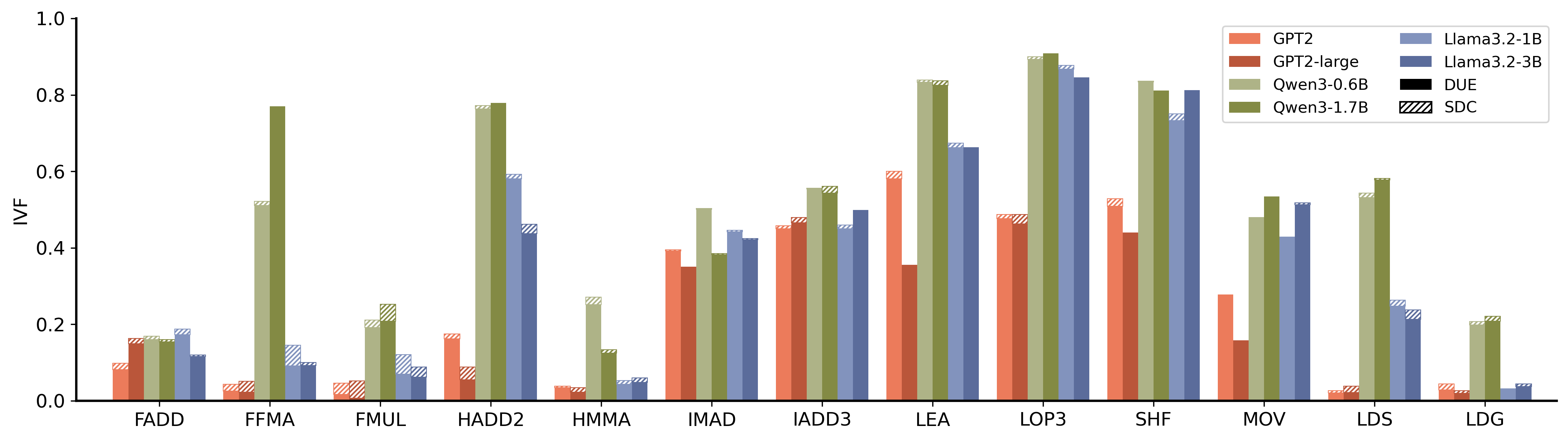}
    \caption{Instruction vulnerability factor (IVF) of different instructions. Solid color represents the portion contributed by DUE and the diagonal line represents the portion contributed by SDC.}
    \label{fig7}
\end{figure*}

\begin{figure}[htbp]
    \centering
    \includegraphics[width=0.46\textwidth]{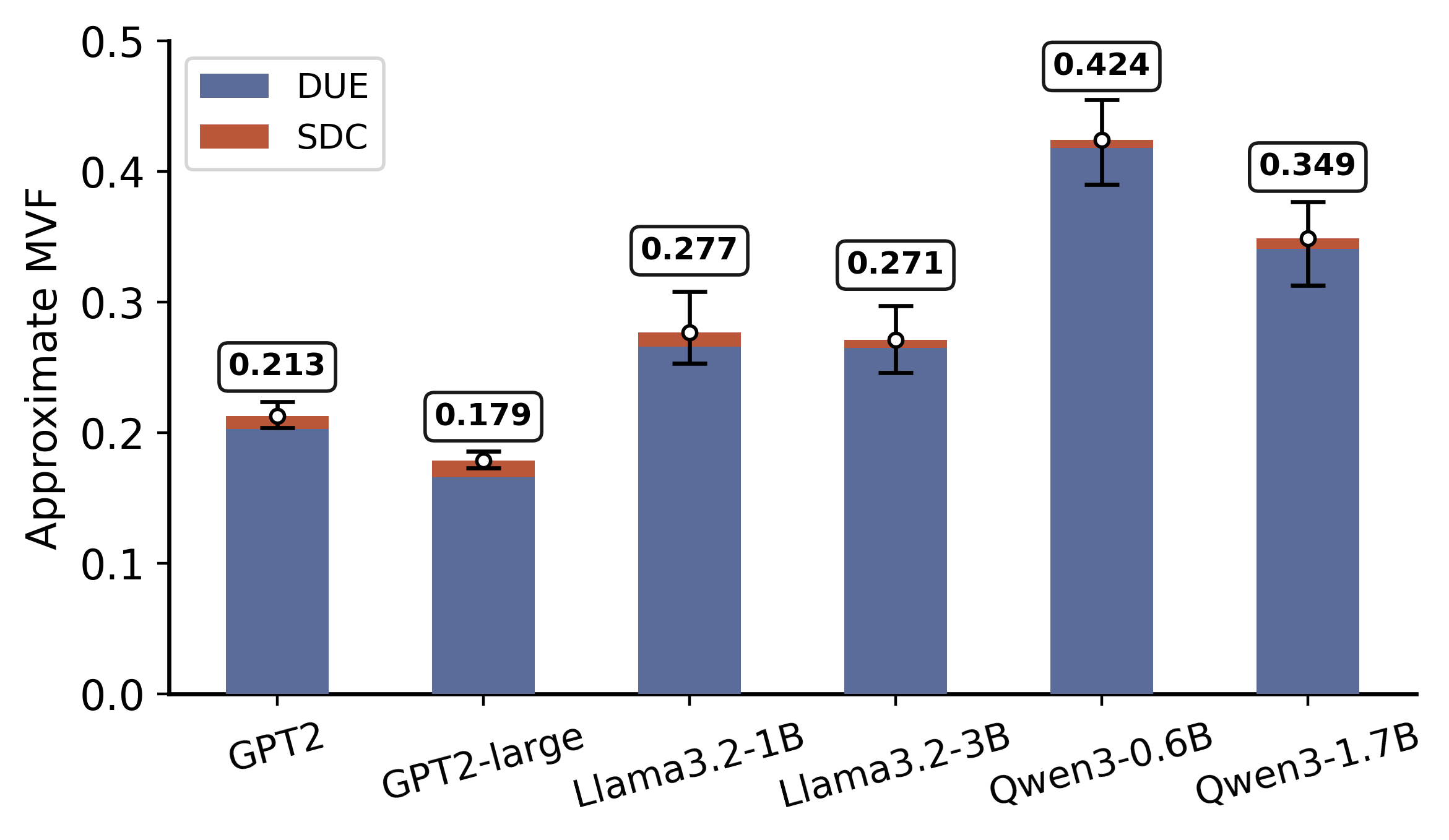}
    \caption{Approximate model vulnerability factor of different models. The error bars represent the standard error of the MVF over multiple experiments.}
    \label{fig8}
\end{figure}

\begin{figure}[htbp]
    \centering
    \includegraphics[width=0.46\textwidth]{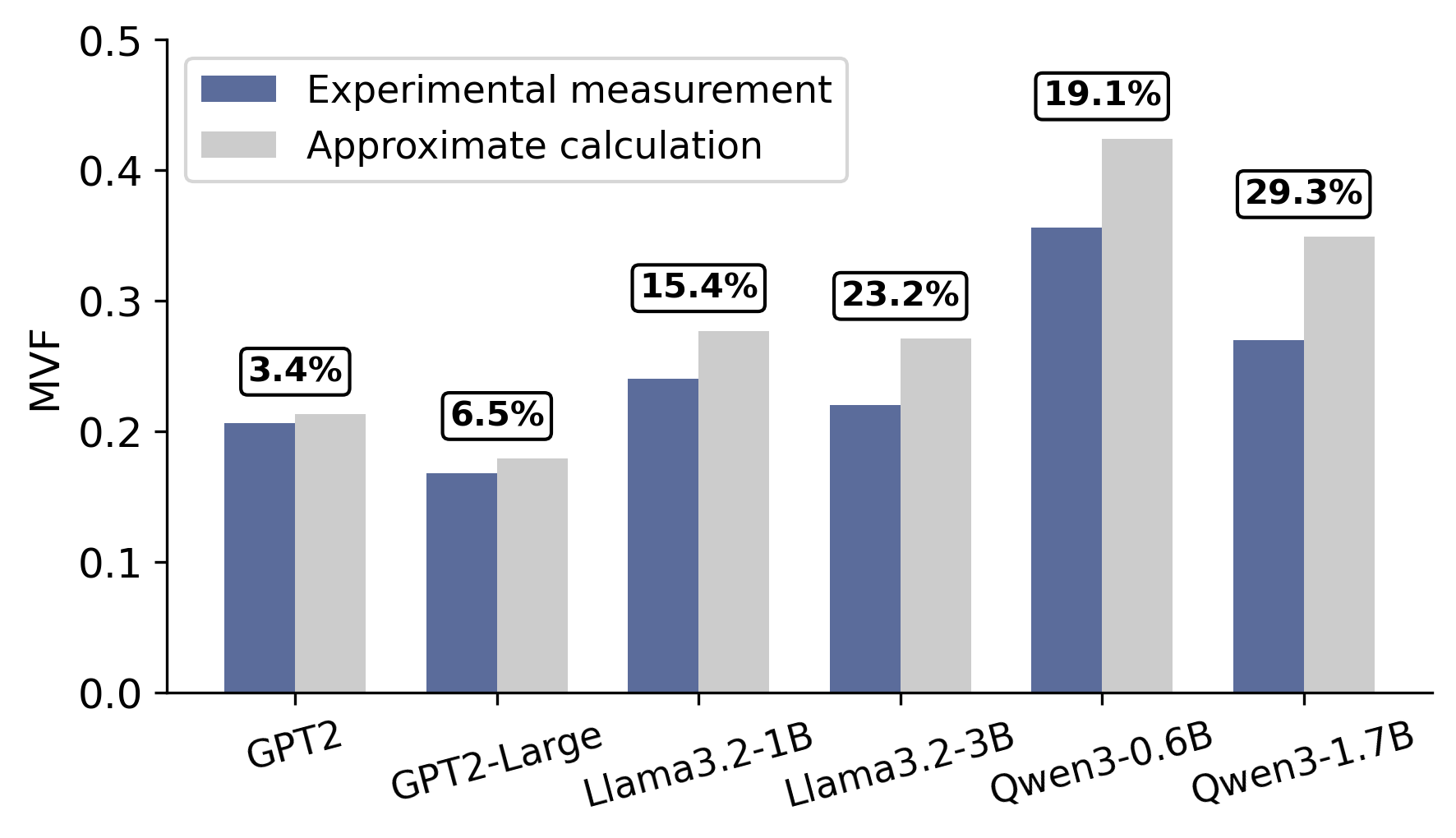}
    \caption{Comparison of MVF obtained from experimental measurements and approximate computations. The numbers on top of the bars indicate the percentage difference.}
    \label{fig9}
\end{figure}

\subsubsection{Approximate Computation of the MVF}
We further explore the instruction vulnerability factor (IVF) with fault injection experiments. The results are presented in Figure~\ref{fig7}. Note that IVF represents the likelihood that an instruction results in faulty outputs of the target LLM in the presence of bit-flip errors. We further leverage the IVF to estimate the MVF, thereby enabling efficient MVF evaluation and validating the experimental results. To this end, we propose a weighted approximation method, where IVFs are aggregated using the proportion of each instruction type as weights, as shown in Algorithm~\ref{alg:mvf}. Some instruction categories contain very few samples, making exhaustive evaluation costly. We therefore further estimate MVF bounds using an approximation strategy that significantly reduces the time. Specifically, for unmeasured instructions within the same group, their IVF is approximated by the average IVF of measured instructions in that group. The resulting aggregated IVFs are used to estimate the MVF, and the upper and lower bounds are computed in a similar manner. The resulting approximated MVF is shown in Figure~\ref{fig8}.

\begin{algorithm}[t]
\caption{Approximate Computation of the MVF}
\label{alg:mvf}
\begin{algorithmic}[1]
\Require \\
Instruction set $I$ \\
Measured IVF $v_i$ for a subset of instructions $M\subseteq I$ \\
Instruction group mapping $g(i)$ \\
Dynamic proportion $p_i$ of each instruction
\Ensure Approximate MVF $V_{avg}$ and its bounds $V_{min}$, $V_{max}$

\For{each instruction $i\in I$}
\If{$i\in M$}
\State $v_i^{avg}\gets v_i$, $v_i^{min}\gets v_i$, $v_i^{max}\gets v_i$
\Else
\State $M_{g(i)}\gets \{j\in M \mid g(j)=g(i)\}$
\State $v_i^{avg}\gets \frac{1}{|M_{g(i)}|}\sum_{j\in M_{g(i)}}v_j$
\State $v_i^{min}\gets \min_{j\in M_{g(i)}}v_j$
\State $v_i^{max}\gets \max_{j\in M_{g(i)}}v_j$
\EndIf
\EndFor

\For{each $s\in\{\textit{avg},\textit{min},\textit{max}\}$}
\State $V_s \gets \sum_{i\in I} p_i v_i^s$
\EndFor

\State \Return $V_{avg}, V_{min}, V_{max}$
\end{algorithmic}
\end{algorithm}

We compare the MVF shown in Figure~\ref{fig6} with the approximated MVF in Figure~\ref{fig8} to evaluate the effectiveness of the proposed method. As shown in Figure~\ref{fig9}, the two approaches exhibit highly consistent MVF trends. Although the approximation introduces numerical errors, the deviation remains below 24\% for all models except Qwen3-1.7B (29.3\%), and is below 7\% for GPT2. We attribute these deviations primarily to the correlations and computational dependencies among instructions, which violate the independence assumption underlying the approximation. Consequently, the approximated MVF tends to overestimate the overall model vulnerability. Nevertheless, it accurately captures the relative vulnerability trends across different models, demonstrating its effectiveness for comparative vulnerability analysis.

\subsection{Instruction-Level}
\subsubsection{Instruction Types}
\begin{figure*}[htbp]
    \centering
    \includegraphics[width=0.92\textwidth]{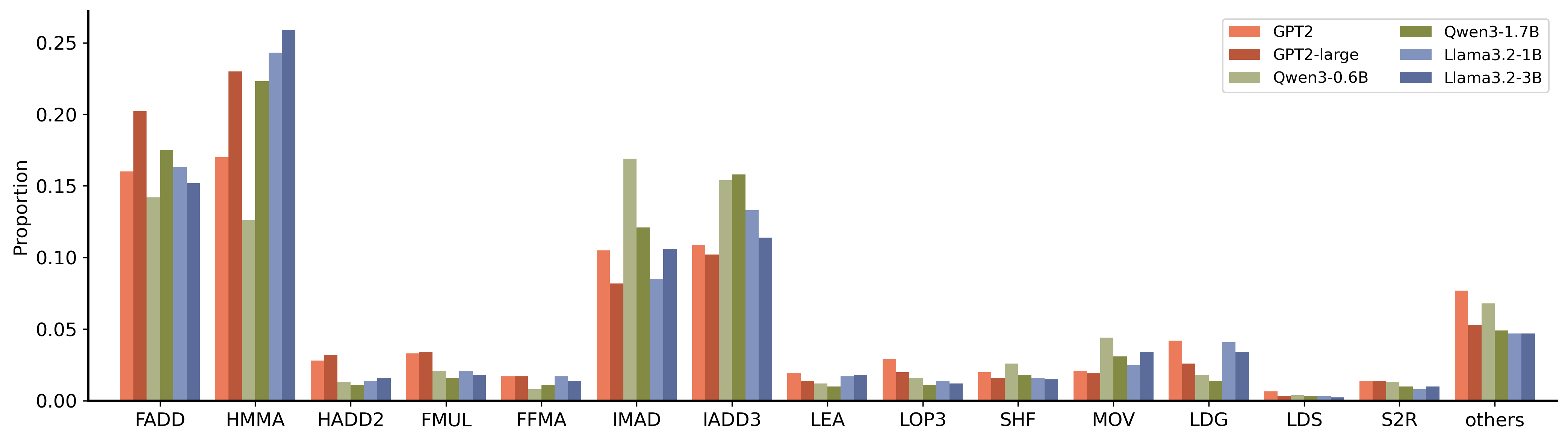}
    \caption{Instruction distribution of different LLMs. Instructions that are fault-injectable but have a low proportion (typically less than 0.01) are grouped into the Others category.}
    \label{fig10}
\end{figure*}

We analyze the distribution of instruction types in different LLMs. As shown in Figure~\ref{fig10}, the computation instructions account for the largest proportion of LLMs, averaging 71.8\%, mainly consisting of FADD, HMMA, IMAD, and IADD3. This dominance is obvious, since LLMs inherently involve intensive data computation. Data transfer instructions, such as MOV and load/store instructions, account for an average of 5.8\%. 

In addition, as shown in Figure~\ref{fig7}, different instruction types exhibit significant differences in IVF. Instructions such as FADD, FMUL, HMMA, and LDG are relatively resilient, whereas SHF, LOP3, and LEA are considerably more vulnerable. This is mainly because LOP3 and LEA are often associated with address computations, where errors in the address calculation chain can easily lead to illegal memory accesses or incorrect data loading. Fortunately, highly vulnerable instructions such as SHF, LEA, and LOP3 occur infrequently in LLM inference, while frequently executed instructions such as FADD and HMMA are among the least vulnerable.

We further observe that different instruction categories tend to induce different error outcomes. Integer computation instructions primarily lead to DUEs across all models, as they are heavily involved in address calculations. In contrast, floating-point instructions are more likely to cause SDCs, accounting for 59\% of all SDCs, especially in the GPT2 and Llama3.2 families. This suggests that protecting floating-point computation instructions is important for mitigating SDCs, whereas address-related integer instructions are more critical for reducing DUEs.

Across architectures, the IVF of different instruction types shows both consistency and variability. Instructions such as FADD, IMAD, and IADD3 exhibit relatively stable vulnerability characteristics across LLM architectures, while LDS, LDG, FFMA, and HADD2 show noticeable architecture-dependent variations. These variations may result from differences in computation patterns, memory access behavior, and kernel implementations across LLM architectures. As a result, architecture-dependent instruction sensitivity becomes an important instruction-level source of the cross-architecture differences observed in MVF.

\begin{tcolorbox}[colback=gray!20,
  colframe=gray!20, 
  boxrule=0pt, 
  left=1mm,
  right=1mm,
  top=1mm,
  bottom=1mm]
{\textbf{Insight \#1:}} Instruction vulnerability in LLMs is primarily determined by instruction type rather than frequency. These vulnerability patterns remain largely consistent across models and architectures, with only a few instruction types exhibiting architecture-dependent variations.
\end{tcolorbox}

\subsubsection{Bit Positions}
\begin{figure*}[htbp]
    \centering
    \includegraphics[width=0.96\textwidth]{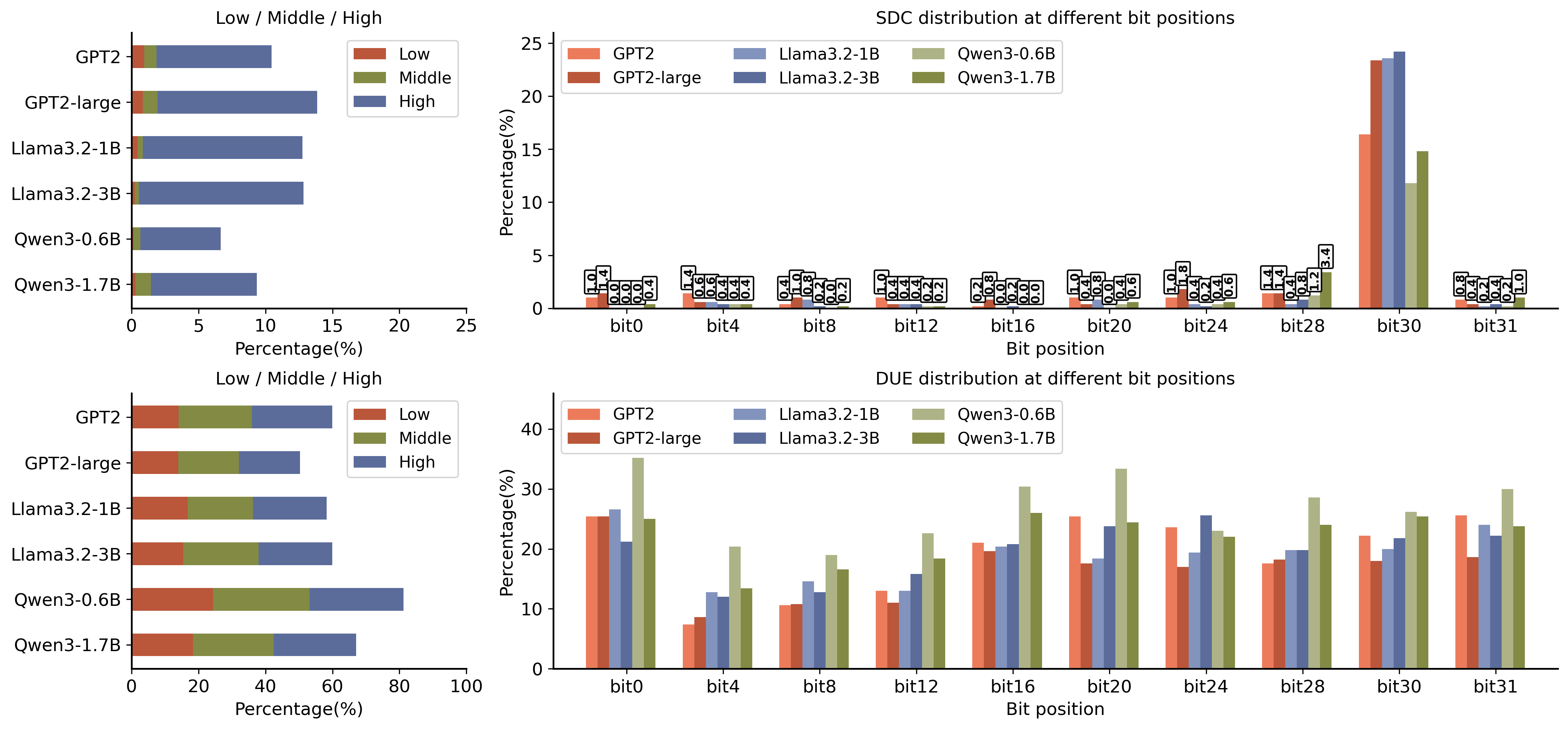}
    \caption{The impact of bit-flip errors at different bit positions on system reliability in terms of DUE and SDC.}
    \label{fig11}
\end{figure*}

To investigate the vulnerability characteristics of models at different bit positions, using the framework introduced in Section~III, we conduct the following experiment by explicitly specifying the target bit position for fault injection. For each bit position, 500 fault injection runs are performed. To reduce the overall runtime, we inject faults every 4 bits from bit 0 to bit 28, and additionally include bits 30 and 31. The results are shown in Figure~\ref{fig11}, which illustrates the impact of bit flips at different positions on LLM reliability, quantified in terms of both DUE and SDC. Several key observations can be made as follows.

In the low-order bits range (bit 0–bit 12), the impact of faults is relatively small. The DUE probability mostly lies between 10\% and 20\%, with an average of 17.15\%. The SDC rate remains below 1\% for all models, and even below 0.5\% for Llama3.2 and Qwen3, with the minimum as low as 0.15\%. This indicates that bit flips in low-order bits are almost harmless and have limited impact on the overall model accuracy, suggesting that little fault-tolerance overhead is needed for these bits. However, bit 0 exhibits a higher DUE rate, with an average of 26.4\%, and thus requires special handling and additional protection.

In the middle-order bits range (bit 16–bit 28), the DUE probability increases on average by 5.34\%, reaching 17\%–33.4\%. The effect is more pronounced in smaller models such as GPT2 and Qwen3-0.6B, where the average DUE probability on Qwen3-0.6B reaches 28.85\%. In contrast, the SDC rate in this range remains between 0.3\% and 1.15\%, which represents no substantial increase compared with low-order bits. This implies that bit flips in middle-order bits are still unlikely to cause SDC. Therefore, middle-order bit errors mainly cause DUE rather than accuracy degradation.

High-order bits (bit 30–bit 31) are the most sensitive region to soft errors. The DUE probability remains in the 18\%–30\% range, and the SDC probability of bit 30 can reach 23.4\% and 24.2\% on GPT2 and Llama3.2, which are 26.5× and 95.8× higher than those of the corresponding low-order bits. This “bit 30 peak” comes from the IEEE single-precision format: bit 30 is the most significant bit of the exponent, so its flip induces numerical deviations spanning several orders of magnitude that are amplified during forward propagation and therefore are very likely to evolve into SDCs. In contrast, the adjacent bit 31 is merely the sign bit. Its flip only changes the sign and is often masked by subsequent nonlinear and normalization operations, making its contribution to SDCs far smaller than that of bit 30.

\begin{tcolorbox}[colback=gray!20,
  colframe=gray!20, 
  boxrule=0pt, 
  left=1mm,
  right=1mm,
  top=1mm,
  bottom=1mm]
\textbf{Insight \#2:} SDC and DUE exhibit distinct bit-level vulnerability patterns. SDC is dominated by high-order bits, whereas DUE is distributed more uniformly across bit positions, motivating different protection strategies.
\end{tcolorbox}

\subsection{Model-Level}
\subsubsection{Component Types}
In this subsection, we conduct random fault injection experiments using our framework, where a single fault is injected in each run. For each model, we perform 5,000 fault injection runs.

By comparing the intermediate activations at each layer during inference with their corresponding golden values, we identify the component affected by the injected fault and derive its vulnerability. Specifically, the vulnerability of a component is defined as the ratio between the number of erroneous model outputs and the number of cases in which the component output deviates from the corresponding golden output. Since some faults may be masked during inference, only cases with observable deviations in the component output are considered.

\begin{figure}[t]
    \centering
    \includegraphics[width=0.46\textwidth]{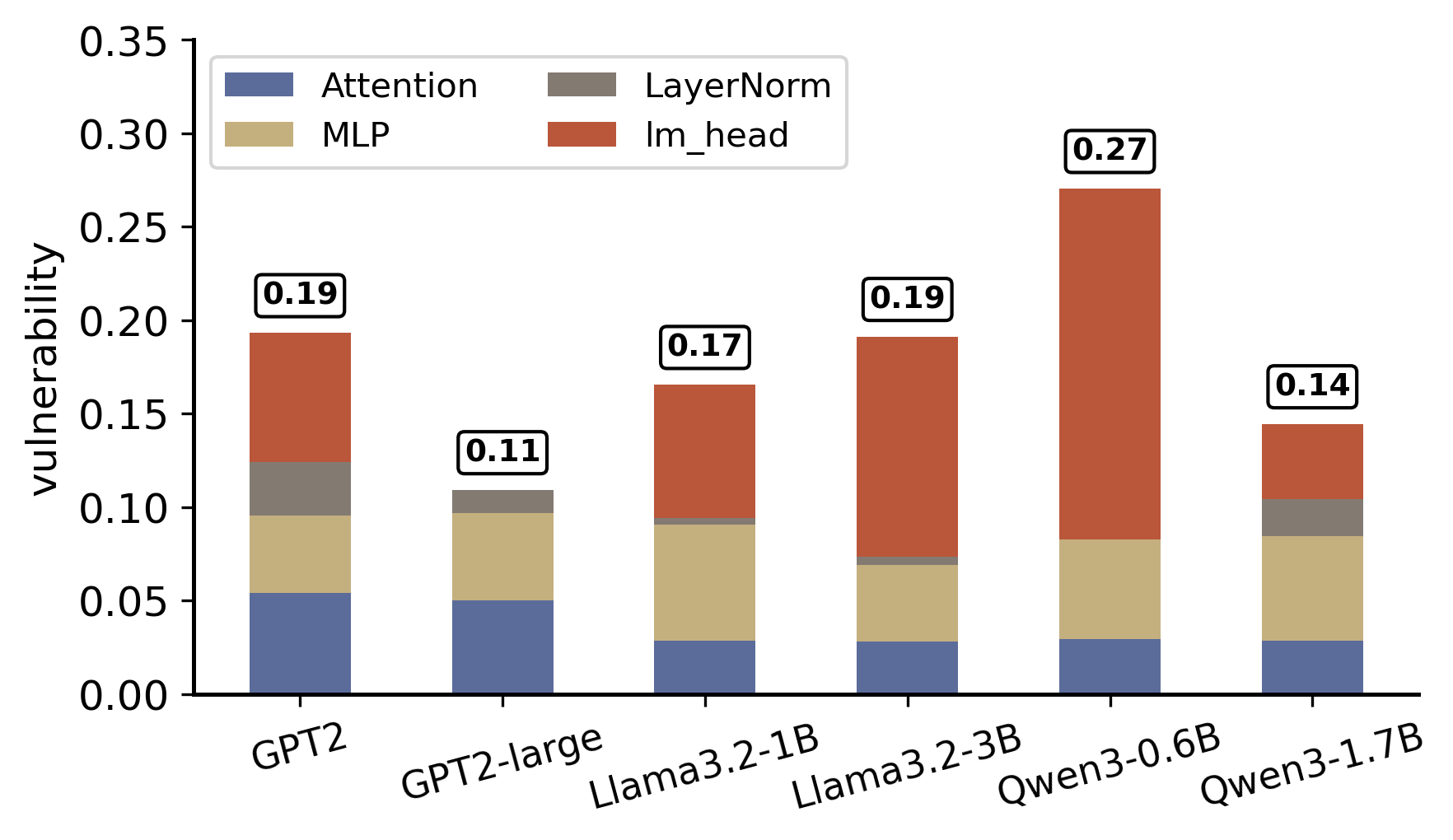}
    \caption{Component vulnerability across LLM architectures.}
    \label{fig12}
\end{figure}

In our experiments, we analyze four main components and quantify their vulnerabilities, as shown in Figure~\ref{fig12}. Overall, these components exhibit markedly different vulnerabilities under soft errors. Among them, $lm\_head$ consistently demonstrates the highest susceptibility across all models. In Qwen3-0.6B, its vulnerability even exceeds 0.18, making it the weakest component and indicating that the output layer plays a particularly critical role in the final prediction of LLMs. In contrast, $LayerNorm$ shows the lowest vulnerability (generally below 0.02), suggesting that normalization operations possess strong fault tolerance. $Attention$ and $MLP$ fall within an intermediate range (approximately 0.03–0.06): they exert a non-negligible influence on overall behavior, but are clearly less critical than the $lm\_head$.

Regarding model scale, different model architectures exhibit distinct trends. For GPT2 and Qwen3, models with more parameters show lower vulnerabilities across all components than their smaller counterparts, with reductions of 43.75\% and 46.26\%, respectively, which are particularly pronounced in the $lm\_head$. In contrast, for Llama3.2, the 3B model exhibits an overall vulnerability 14.63\% higher than the 1B model, with a more pronounced increase in the $lm\_head$.

\begin{tcolorbox}[
  colback=gray!20,
  colframe=gray!20, 
  boxrule=0pt, 
  left=1mm,
  right=1mm,
  top=1mm,
  bottom=1mm]
\textbf{Insight \#3:} Component-level vulnerability in LLMs is highly heterogeneous, with $lm\_head$ showing higher vulnerability and $LayerNorm$ exhibiting stronger robustness to soft errors, motivating component-aware fault protection.
\end{tcolorbox}

\subsubsection{Layer Variations}

\begin{figure*}[htbp]
    \centering
    \includegraphics[width=0.96\textwidth]{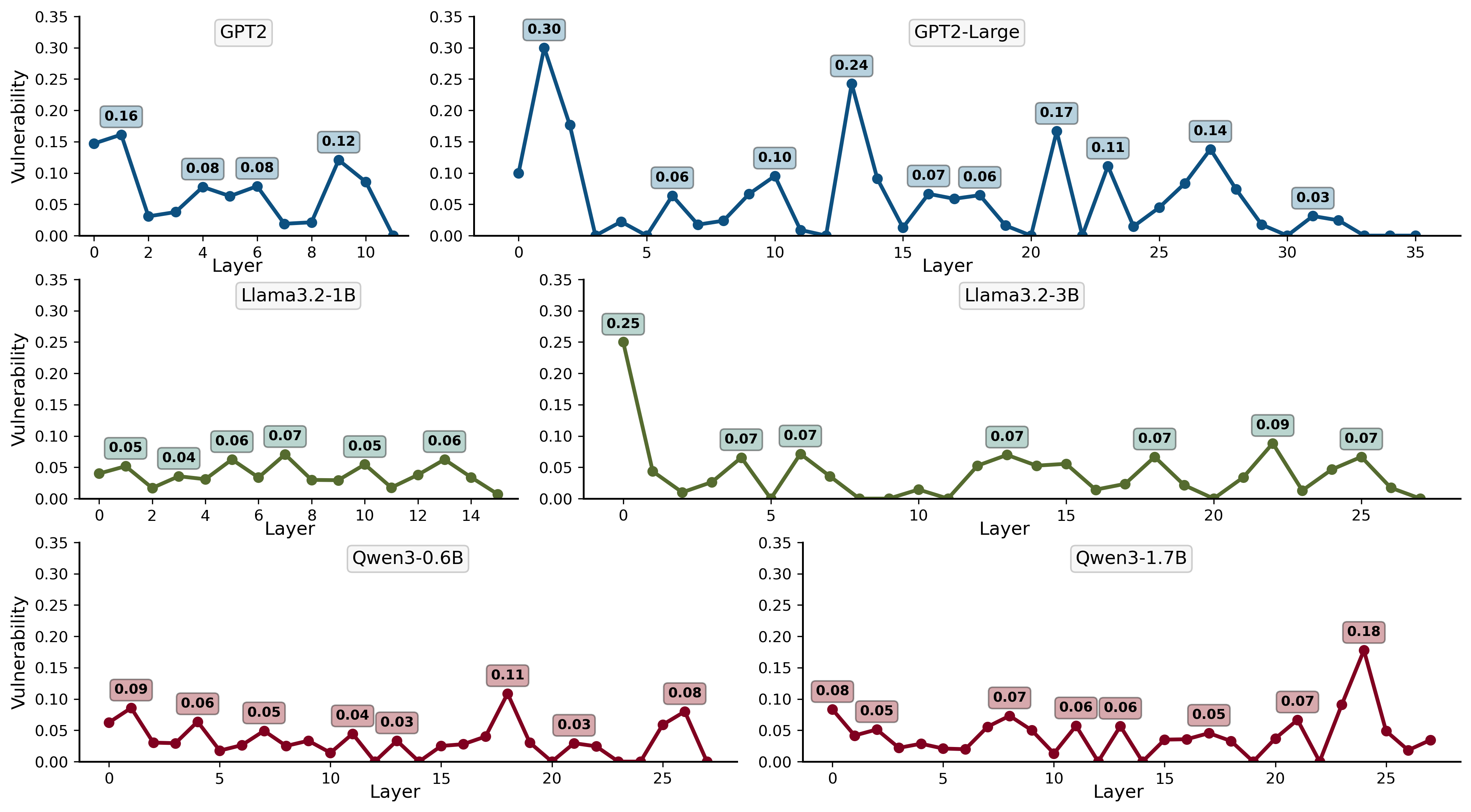}
    \caption{Comparison of error rates in different layers. The horizontal axis represents different layers, and the vertical axis represents vulnerability.}
    \label{fig13}
\end{figure*}

Figure~\ref{fig13} presents the experimental results obtained from 5,000 fault injections per model. The plot illustrates the vulnerability metric across layer indices. Through cross-model comparison, we identify the following key observations.

In the GPT2 family of models, GPT2-large shows a markedly higher overall error rate than GPT2, with vulnerable layers more widely distributed and a maximum error rate close to 0.3. The most vulnerable layers are mainly located at layers 0–2, 13, 21, 23, and 27. In comparison, GPT2 has a lower overall error rate, but several layers (especially layers 0–1 and 9–10) still exhibit pronounced vulnerability. 

In the Llama3.2 family of models, Llama3.2-1B has the lowest overall error rate, with peaks below 0.1 and relatively small variation across layers. In Llama3.2-3B, however, the error rate of layer 0 approaches 0.3, indicating high vulnerability, while the error rates of the middle and deeper layers rapidly decrease and then stabilize. This pattern suggests that, as the parameter scale increases, the first layer of Llama3.2 models becomes significantly more sensitive to soft errors, whereas subsequent layers tend to be more robust. 

In the Qwen3 family of models, overall error rates remain low, with most layers below 0.05. Nonetheless, several locally vulnerable layers still appear, such as layer 18 of Qwen3-0.6B and layer 24 of Qwen3-1.7B. This indicates that, although Qwen3 is generally more robust, a few individual layers can still act as potential high-risk points that warrant special attention in practical deployments.

Moreover, we observe an interesting phenomenon: as the model parameter scale increases, the overall shape of the layer-wise vulnerability distribution remains largely unchanged, while the peak error rates increase significantly.

\begin{tcolorbox}[
colback=gray!20,colframe=gray!20, boxrule=0pt, 
left=1mm,right=1mm,top=1mm, bottom=1mm]
\textbf{Insight \#4:} Model scaling preserves the overall layer-wise vulnerability pattern while amplifying vulnerability at critical layers.
\end{tcolorbox}

\begin{figure*}[t]
    \centering
    \includegraphics[width=0.96\textwidth]{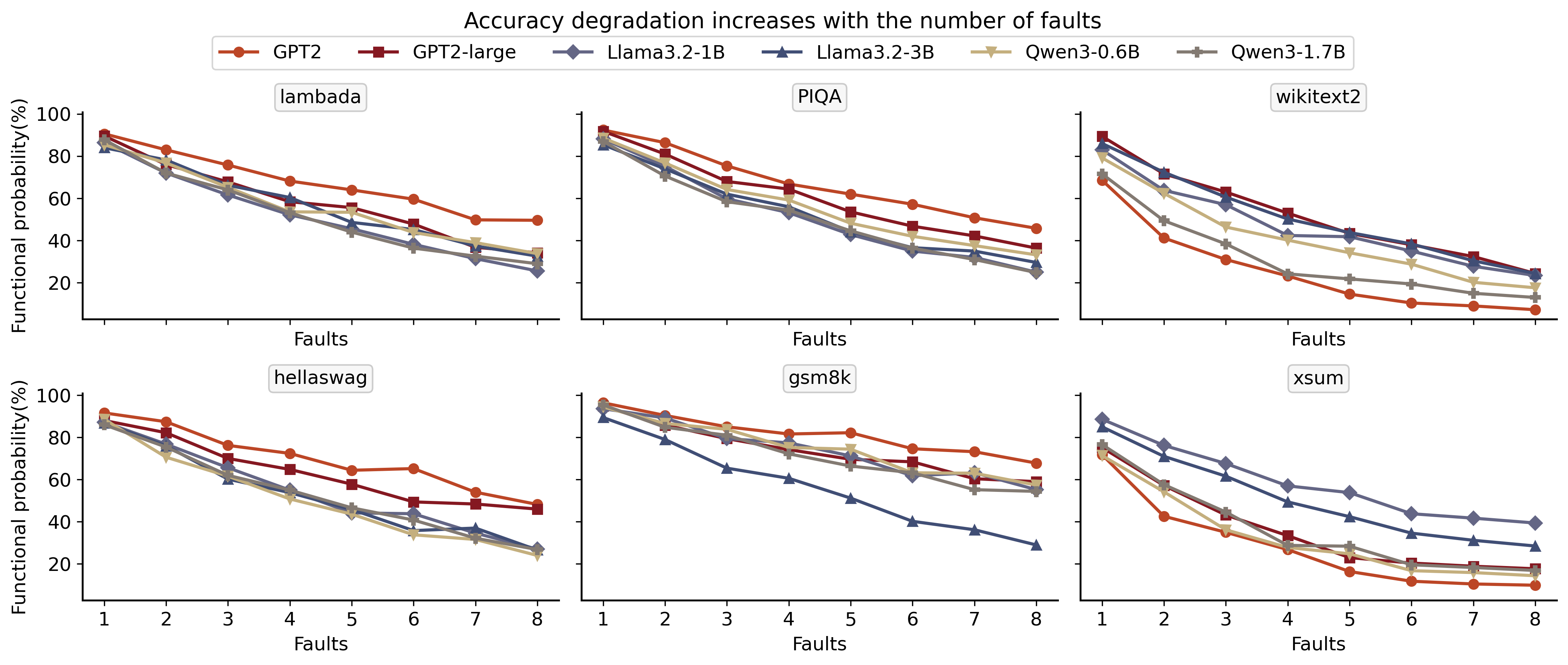}
    \caption{The model’s functional probability decreases as the number of faults injected in a single run increases, where functional probability is defined as the proportion of correct inferences among all experiments.}
    \label{fig14}
\end{figure*}

\begin{table*}[t]
\caption{The scores of the models on different datasets.\label{table4}}
\centering
\renewcommand{\arraystretch}{1.05}
\begin{tabular}{c c c c c c c}
\toprule
Model & Lambada(ACC) & PIQA(ACC) & HellaSwag(ACC) & WikiText-2(PPL) & XSum(ROUGE-1) & GSM8K(ACC)\\
\midrule
GPT2 & 32.56 & 62.89 & 28.92 & 37.37 & 11.66 & 1.14\\
GPT2-large & 47.66 & 70.35 & 36.40 & 22.61 & 11.80 & 1.52\\
Llama3.2-1B & 62.22 & 74.32 & 47.68 & 11.56 & 12.74 & 6.60\\
Llama3.2-3B & 70.04 & 76.71 & 55.21 & 9.26 & 12.86 & 25.85\\
Qwen3-0.6B & 40.07 & 67.30 & 37.47 & 26.14 & 11.85 & 41.70\\
Qwen3-1.7B & 50.30 & 72.09 & 46.14 & 21.04 & 13.06 & 69.14\\
\bottomrule
\end{tabular}
\end{table*}

\subsection{Task-Level}
Previous studies~\cite{xue2026adaptive} have shown that variations in input can lead to changes in model vulnerability. Although a model may appear highly reliable for certain inputs, there can still exist specific cases that significantly degrade its robustness. However, such observations are typically restricted to different samples within the same dataset. Motivated by this, we investigate whether LLMs exhibit different degrees of vulnerability under tasks of varying difficulty. To this end, we design a vulnerability analysis experiment for LLMs under different task difficulties, in order to characterize how model robustness changes across tasks. We select six datasets representing different difficulty levels—Lambada, PIQA, HellaSwag, WikiText-2, XSum, and GSM8K—which cover a broad range of scenarios including text generation, mathematical and commonsense reasoning, and language modeling. The performance of these models on the six datasets is summarized in Table~\ref{table4}, and the difficulty posed by each dataset to the models varies considerably.

For datasets evaluated by accuracy, we choose single samples that can be consistently inferred correctly under fault-free conditions as test cases, since our focus is on whether injected faults cause deviations from the correct outputs. For datasets evaluated by PPL or ROUGE-1, we instead use a single text passage for evaluation. Because NVBitFI requires the execution of thousands of CUDA kernels during LLM inference, the time overhead is extremely high. In our experiments, running XSum on just one text passage generated trillions of dynamic instructions, making full-dataset fault injection prohibitively time-consuming. Under a fully random fault injection setting, we conduct 4,000 fault injection runs for each model.

Our primary focus is the degree of functional probability degradation caused by faults. As illustrated in Figure~\ref{fig14}, the LLMs exhibit comparable levels of degradation on Lambada, PIQA, HellaSwag, and GSM8K, whereas the degradation is more pronounced on WikiText-2 and XSum. Moreover, as the number of injected faults increases, the functional probability of all models shows a clear downward trend across all datasets, but the magnitude and shape of the degradation curves vary substantially. We conduct a comparative analysis from the following three perspectives:

\begin{table}[t]
\centering
\caption{The reliability changes of models with the same architecture after increasing the number of parameters.
\label{table5} \\
($\uparrow$ represents more reliability, $\downarrow$ represents less reliability).}
\renewcommand{\arraystretch}{1.05}
\setlength{\tabcolsep}{3pt}
\begin{tabular}{c c c c c c c}
\toprule
Model & Lambada & PIQA & HellaSwag & WikiText-2 & XSum & GSM8K \\
\midrule
GPT2 & $\downarrow$ & $\downarrow$ & $\downarrow$ & $\uparrow$ & $\uparrow$ & $\downarrow$\\
Llama3.2 & $\uparrow$ & $\uparrow$ & $\downarrow$ & $\uparrow$ & $\downarrow$ & $\downarrow$\\
Qwen3 & $\downarrow$ & $\downarrow$ & $\uparrow$ & $\downarrow$ & $\uparrow$ & $\downarrow$\\
\bottomrule
\end{tabular}
\end{table}

\subsubsection{Differences across tasks}
Among the six tasks, GSM8K shows relatively strong robustness, with some models still maintaining a functional probability of approximately 60\% even after eight fault injections. In contrast, WikiText-2 and XSum suffer the steepest declines. For example, GPT2 on WikiText-2 displays an almost cliff-like drop, with functional probability plunging from 68.6\% to below 10\%. This suggests that tasks centered on language modeling and long-text generation are more vulnerable to perturbations, whereas reasoning-oriented (GSM8K) and cloze-style (Lambada) tasks are comparatively less sensitive.

\subsubsection{Differences across architecture}
In terms of architecture, GPT2 consistently demonstrates stronger robustness than Llama3.2 and Qwen3. For instance, on PIQA, GPT2-large still retains about 36.4\% functional probability after eight rounds of fault injection, whereas Llama3.2-1B has already dropped to 25.0\%. Similarly, GPT2 shows a slower degradation curve on Lambada and GSM8K. In contrast, the Qwen models exhibit moderate behavior: although their initial accuracy is relatively high, they degrade more rapidly under fault perturbations, especially on WikiText-2 and XSum.

\subsubsection{Impact of parameter scaling}
To examine the effect of parameter scaling on fault robustness, we summarize the results in Table~\ref{table5}. Scaling effects are clearly task- and architecture-dependent. For Lambada and PIQA, larger Llama3.2 models improve reliability, while GPT2 and Qwen3 degrade. In XSum and WikiText-2, benefits vary across architectures, indicating strong task sensitivity. On challenging reasoning tasks like HellaSwag and GSM8K, most scaled models suffer larger accuracy drops, suggesting parameter growth alone does not ensure fault robustness. Although larger models often achieve better fault-free performance, they may degrade faster under faults, as seen in GPT2-large and Llama3.2-3B. Thus, increasing parameter scale improves representational capacity but may amplify fault-induced degradation.

\begin{tcolorbox}[colback=gray!20,
  colframe=gray!20, 
  boxrule=0pt, 
  left=1mm,
  right=1mm,
  top=1mm,
  bottom=1mm]
\textbf{Insight \#5:} Fault robustness in LLMs is highly task- and architecture-dependent, and the impact of model scaling varies across tasks and architectures.
\end{tcolorbox}

\section{Conclusion}
We conduct a large-scale instruction-level fault injection study on GPU-based LLM inference systems, aiming to systematically characterize the manifestations and root causes of SDCs and DUEs during inference. Through extensive fault injection experiments, we quantify the instruction vulnerability factor across different instruction types and leverage these measurements to approximate the overall instruction-level vulnerability of different LLMs. These results provide quantitative, instruction-level evidence for assessing and comparing the reliability of LLM inference systems. In addition, we conduct a comprehensive analysis of the factors affecting LLM vulnerability across three abstraction levels and five dimensions, leading to five key empirical observations on the abnormal behavior of LLM inference.
% \nocite{*}

\section*{Acknowledgments}
The authors used GPT-5 to polish the manuscript’s text. All core ideas, methodologies, and scientific conclusions are entirely the original work of the authors, who have thoroughly reviewed all AI-assisted outputs and take full responsibility for the final content.

\bibliographystyle{IEEEtran}
\bibliography{ref}

\end{document}